# Derivation of Bose-Einstein and Fermi-Dirac statistics from quantum mechanics: Gauge-theoretical structure

Yuho Yokoi[1] and Sumiyoshi Abe[2,3,4 *)]

[1] Graduate School of Engineering, Mie University, Mie 514-8507, Japan
[2] Physics Division, College of Information Science and Engineering,
  Huaqiao University, Xiamen 361021, China
[3] Department of Physical Engineering, Mie University, Mie 514-8507, Japan
[4] Institute of Physics, Kazan Federal University, Kazan 420008, Russia

**Abstract.**  A possible quantum-mechanical origin of statistical mechanics is discussed, and microcanonical and canonical ensembles of bosons and fermions are derived from the stationary Schrödinger equation in a unified manner. The interaction Hamiltonians are constructed by the use of the discrete phase operators and the gauge-theoretical structure associated with them. It is shown how the interaction Hamiltonians stipulated by the gauge symmetry generate the specific patterns of entanglement that are desired for establishing microcanonical ensembles. A discussion is also made about the interrelation between random phases and perfect decoherence in the vanishing-interaction limit.

___________

*) Corresponding author

At the fundamental level of its principles, statistical mechanics may be inseparable from quantum mechanics. An evidence for this may be seen in the fact that the Planck constant is indispensable even for formulating classical statistical mechanics in the phase space $\Gamma$. In addition, the probabilistic concept in classical theory is due to lack of knowledge or information about dynamical details of a large number of particles, whereas in quantum mechanics it is an element of the laws of nature. It is also worth recalling that the quantum-classical correspondence in statistical mechanics is concerned not with the Planck constant but with the temperature: classical statistical mechanics appears in the high-temperature regime of quantum statistical mechanics.

In recent years, a lot of works have been done on understanding emergence of statistical mechanics from quantum mechanics. Relatively earlier attempts are found in [1-3], and then a crucial role of quantum entanglement [4,5], ensemble typicality [6,7] and eigenstate thermalization [8,9] (and the references cited therein) have come to form integral parts of this subject. These are also of direct relevance to quantum thermodynamics [10,11] that provides nanoscience and nanotechnology with basic theoretical tools.

To derive statistical mechanics, i.e., Bose-Einstein and Fermi-Dirac statistics, from quantum mechanics, we require the following two points: (I) perfect decoherence has to be realized for an isolated system, and (II) the principle of equal *a priori* probability should be understood through a solution of the Schrödinger equation. Once these requirements are simultaneously fulfilled, then it will be possible to construct a microcanonical ensemble of the isolated system.

Some comments are in order. Firstly, derivations of microcanonical and canonical ensembles of isolated fermions have been discussed in Ref. [12], where the four-body interaction is assumed, and a "chaotic eigenstate", which may allow to express expectation values and variances of physical quantities only in terms of the occupation number in each single-particle state, is considered. The averaging procedure analogous to the microcanonical one is a basic premise there. In addition, requirement (I), i.e., realization of perfect decoherence of the state, is not investigated. Secondly, "eigenstate typicality" has recently been studied also for free fermions in Ref. [13]. There, only the states of a subsystem are discussed through partial trace over the *complement system* (which would remind us of the tildian system in Takahashi-Umezawa thermofield dynamics [14]) that are entangled with the objective fermionic subsystem. Therefore, the discussion is about canonical ensemble, not microcanonical ensemble. Our subsequent discussion will be in marked contrast to these works.

For the purpose of simultaneously establishing (I) and (II), clearly a certain guiding principle is needed for determining an interaction Hamiltonian. Interaction is certainly indispensable even for statistical mechanics of *free* particles. As known in the discussion of the classical ideal gas, actually the particles should weakly interact each other: otherwise, the equilibrium state may not be realized. However, such an interaction is ignored at the juncture when description of the system shifts to be statistical mechanical. This philosophy is respected in the present work.

Here, we study interactions of the specific type for both bosons and fermions. They have a gauge-theoretical origin, induce quantum entanglement of the desired forms in the solutions of the stationary Schrödinger equations and are required to vanish when the statistical mechanical descriptions of the systems are performed. We discuss how disappearance of the interactions is linked with random phases that can give rise to prefect decoherence of the quantum states of the isolated systems. Microcanonical ensembles of bosons and fermions are found to emerge, and then canonical ensembles are further constructed from them.

Let us start our discussion with considering the following free Hamiltonian of $N$ *identical* oscillators:

$$\hat{H}_0 = \sum_{i=1}^{N} \hat{h}_i, \qquad \hat{h}_i = \varepsilon\, \hat{n}_i. \tag{1}$$

Here, $\hat{n}_i$ is the number operator given by $\hat{n}_i = \hat{a}_i^\dagger\, \hat{a}_i$, where $\hat{a}_i^\dagger$ and $\hat{a}_i$ are respectively the creation and annihilation operators satisfying the commutation relations, $[\hat{a}_i, \hat{a}_j^\dagger] = \delta_{ij}$ and $[\hat{a}_i, \hat{a}_j] = [\hat{a}_i^\dagger, \hat{a}_j^\dagger] = 0$ for bosons, and the anticommutation relations, $\{\hat{a}_i, \hat{a}_j^\dagger\} = \delta_{ij}$ and $\{\hat{a}_i, \hat{a}_j\} = \{\hat{a}_i^\dagger, \hat{a}_j^\dagger\} = 0$ for fermions. $\varepsilon$ denotes the common energy of the identical oscillators, and zero-point energies are not included for the sake of simplicity. In addition, the following notation is understood: $\sum_{i=1}^{N} \hat{A}_i = \hat{A}_1 \otimes \hat{I}_2 \otimes \cdots \otimes \hat{I}_N + \hat{I}_1 \otimes \hat{A}_2 \otimes \hat{I}_3 \otimes \cdots \otimes \hat{I}_N + \cdots + \hat{I}_1 \otimes \cdots \otimes \hat{I}_{N-1} \otimes \hat{A}_N$ with $\hat{I}_i$ being the identity operator in the space of the $i$-th oscillator.

As mentioned above, for formulating statistical mechanics of free particles, it is necessary to introduce interactions, which are ignored at the stage of describing the system in the statistical mechanical manner. The interactions we consider here are closely related to the unitary phase operator defined as follows [15,16]:

$$\exp(i\hat{\phi}_i) = \sum_{m_i=0}^{s} \exp(i\theta_{m_i}) |\theta_{m_i}\rangle_i {}_i\langle\theta_{m_i}| \qquad (i=1,2,...,N), \qquad (2)$$

where $|\theta_{m_i}\rangle_i$ is the discrete phase state defined by

$$|\theta_{m_i}\rangle_i = \frac{1}{\sqrt{s+1}} \sum_{n_i=0}^{s} \exp(i n_i \theta_{m_i}) |n_i\rangle_i. \qquad (3)$$

The quantities and symbols appearing in these expressions are defined as follows. In the case of bosons, $s$ is a large but finite positive integer: that is, each boson is defined in the $(s+1)$-dimensional *truncated* space, and the limit $s \to \infty$ should be taken after all calculations concerning the phases and phase operators [15,16]. On the other hand, $s$ is always *unity* for fermions, and the limiting procedure is irrelevant. $\theta_{m_i}$ is a $c$-number given by

$$\theta_{m_i} = \frac{2\pi m_i}{s+1} \qquad (m_i = 0, ..., s). \qquad (4)$$

$|n_i\rangle_i$ in Eq. (3) is the orthonormal number state satisfying $\hat{n}_i |n_i\rangle_i = n_i |n_i\rangle_i$, and the set $\{|n_i\rangle\}_{n_i=0,...,s}$ forms a complete orthonormal system in the $(s+1)$–dimensional space. Then, it is straightforward to ascertain that the set $\{|\theta_{m_i}\rangle\}_{m_i=0,...,s}$ also forms a complete orthonormal system in the same space. Therefore, the state in Eq. (2) is the eigenstate of

the operator in Eq. (1) with the eigenvalue $\exp(i\theta_{m_i})$. The total number state is given by $|n_1, n_2, ..., n_N\rangle = (\hat{a}_1^\dagger)^{n_1} (\hat{a}_2^\dagger)^{n_2} \cdots (\hat{a}_N^\dagger)^{n_N} |0\rangle / \sqrt{n_1! n_2! \cdots n_N!}$ ($n_i = 0, ..., s$; $i = 1, 2, ..., N$), where $|0\rangle = \otimes_{i=1}^{N} |0\rangle_i$ with $|0\rangle_i$ being the ground state of the $i$-th oscillator annihilated by $\hat{a}_i$.

The phase states and phase operators mentioned above have widely been applied to the studies of phase properties of quantum states of light such as the coherent and squeezed states.

As pointed out in Ref. [17], the operator $\hat{\phi}_i$ in Eq. (2) can be interpreted as an Abelian gauge field. To see it succinctly, we note the following relations:

$$\exp(i\hat{\phi}_i)|n_i\rangle_i = |n_i - 1\rangle_i \quad (n_i \neq 0), \qquad \exp(i\hat{\phi}_i)|0\rangle_i = |s\rangle_i, \tag{5}$$

which can immediately be obtained from the phase operator expressed in terms of the number states

$$\exp(i\hat{\phi}_i) = \sum_{n_i=0}^{s-1} |n_i\rangle_i {}_i\langle n_i + 1| + |s\rangle_i {}_i\langle 0|. \tag{6}$$

With this form, we observe that under the gauge transformation of the first kind

$$|n_i\rangle_i \to |n_i\rangle_i \exp(i\Lambda_{n_i, \mu_i}), \tag{7}$$

with the local ($n_i$ – dependent) gauge function

$$\Lambda_{n_i,\mu_i} = n_i \theta_{\mu_i} = \frac{2\pi \mu_i n_i}{s+1} \qquad (\mu_i = 0,...,s), \tag{8}$$

$\hat{\phi}_i$ undergoes the following gauge transformation of the second kind:

$$\hat{\phi}_i \rightarrow \hat{\phi}_i - \partial \Lambda_{n_i,\mu_i}, \tag{9}$$

where $\partial \Lambda_{n_i,\mu_i} \equiv \Lambda_{n_i+1,\mu_i} - \Lambda_{n_i,\mu_i}\ (=\theta_{\mu_i})$. Thus, we see that Eq. (5) remains gauge invariant.

Upon constructing the interaction Hamiltonians, we employ this gauge-theoretical structure as the guiding principle. Let us consider the following operator:

$$\hat{V} = \sum_{i=1}^{N} \hat{v}_i, \tag{10}$$

where

$$\hat{v}_i = \left[\exp(i\hat{\phi}_i) - |s\rangle_{i\,i}\langle 0|\right]\exp(-i\theta_{m_i}) \qquad \text{(bosons)}, \tag{11}$$

$$\hat{v}_i = \left[\exp(i\hat{\phi}_i) - |1\rangle_{i\,i}\langle 0|\right]\exp(-i\theta_{m_i})(-1)^{\hat{F}_i} \qquad \text{(fermions)}. \tag{12}$$

The operator $\hat{F}_i$ in Eq. (12) is defined by

$$\hat{F}_i = \sum_{j<i} \hat{n}_j \qquad (\hat{F}_1 \equiv 0). \tag{13}$$

The summations of the operators in Eqs. (10) and (13) are understood in the sense mentioned after Eq. (1). The subtraction terms inside the square brackets on the right-hand sides in Eqs. (11) and (12) are introduced in order to eliminate the second term on the right-hand side in Eq. (6). The factor $(-1)^{\hat{F}_i}$ is characteristic for the fermions. Recall that $|0\rangle_{i\,i}\langle 0|$ commute with any operator, and $|0\rangle_{i\,i}\langle 1| = |0\rangle_{i\,i}\langle 0|\hat{a}_i$ and $\hat{a}_j^\dagger$ ($i \ne j$) anticommute with each other. Therefore, the above-mentioned factor allows us to move the location of $\hat{v}_i$ on the total fermionic number state without sign changes. For example, $\hat{v}_i|n_1, n_2, ..., n_N\rangle = (\hat{a}_1^\dagger)^{n_1}(\hat{a}_2^\dagger)^{n_2}\cdots(\hat{a}_{i-1}^\dagger)^{n_{i-1}}\hat{v}_i(\hat{a}_i^\dagger)^{n_i}\cdots(\hat{a}_N^\dagger)^{n_N}|0\rangle$.

The interaction Hamiltonian we consider here is now given as follows:

$$H_\mathrm{I} = g\left(V^\dagger V + N\sum_{i=1}^{N}|0\rangle_{i\,i}\langle 0|\right), \tag{14}$$

where $g$ is a coupling constant. The second term inside the brackets is related to the subtraction terms in Eqs. (11) and (12). Accordingly, the total Hamiltonian reads

$$\hat{H} = \hat{H}_0 + \hat{H}_\mathrm{I}, \tag{15}$$

which is defined in the $(s+1)^N$ – dimensional space.

We consider the following stationary Schrödinger equation:

$$\hat{H}|u_E\rangle = E|u_E\rangle. \tag{16}$$

The exact solution of this equation is found to be

$$E_{M,N} = M\varepsilon + g f(N), \tag{17}$$

$$|u_E\rangle = |M; N, [\theta_m]\rangle$$
$$= \frac{1}{\sqrt{W(M,N)}} \sum_{P\{n\}} |n_1, n_2, ..., n_N\rangle \delta_{n_1+n_2+\cdots+n_N, M} \exp\left(i\sum_{i=1}^{N} n_i \theta_{m_i}\right). \tag{18}$$

Kronecker's delta inside the summation in Eq. (18) implies the constraint condition that the number of excitation is fixed to be $M$. Clearly, the maximum value of $M$ is $Ns$. $f(N)$ in Eq. (17) has the following properties. $f(N) = N^2$ for the bosons. On the other hand, it varies as $0 < f(N) \leq N^2$ for the fermions, depending on the value of $M$, and the maximum value $f(N) = N^2$ is realized in the case $M = 1$. The symbol $\Sigma_{P\{n\}}$ in Eq. (18) denotes the summation over all independent permutations of $(n_1, n_2, ..., n_N)$. Thus, entanglement of the specific type is induced by the interaction Hamiltonian. For example, in the case of $M = 2$ and $N = 3$, the states are $|2; 3, [\theta_m]\rangle = \frac{1}{\sqrt{6}}$

$$\times \left[ |2,0,0\rangle \exp(2i\theta_{m_1}) + |0,2,0\rangle \exp(2i\theta_{m_2}) + |0,0,2\rangle \exp(2i\theta_{m_3}) + |1,1,0\rangle \exp(i\theta_{m_1} + i\theta_{m_2}) \right.$$

$$\left. + |1,0,1\rangle \exp(i\theta_{m_1} + i\theta_{m_3}) + |0,1,1\rangle \exp(i\theta_{m_2} + i\theta_{m_3}) \right] \quad \text{and} \quad |2; 3, [\theta_m]\rangle = \frac{1}{\sqrt{3}}$$

$$\times \left[ |1,1,0\rangle \exp(i\theta_{m_1} + i\theta_{m_2}) + |1,0,1\rangle \exp(i\theta_{m_1} + i\theta_{m_3}) + |0,1,1\rangle \exp(i\theta_{m_2} + i\theta_{m_3}) \right] \quad \text{for the}$$

bosons and fermions, respectively.

There are two important points here. The first one is concerned with the gauge transformation of the Hamiltonian. From Eqs. (7) and (8), both of $\hat{v}_i$ in Eqs. (11) and (12) transform as $\hat{v}_i \rightarrow \hat{v}_i \exp(-i\theta_{\mu_i})$. Therefore, unlike $\hat{H}_0$, the interaction

Hamiltonian in Eq. (14) is not invariant under the transformation. However, the state in Eq. (18) transforms as $|M; N, [\theta_m]\rangle \rightarrow |M; N, [\theta_m + \theta_\mu]\rangle$. Consequently, Eq. (16) takes the same form for the transformed Hamiltonian and the state (with the energy eigenvalue being kept unchanged), implying gauge covariance. The second point is that the state $|M; N, [\theta_m]\rangle$ in Eq. (18) satisfies the normalization condition

$$\frac{1}{(s+1)^N} \sum_{m_1, m_2, \ldots, m_N = 0}^{s} \langle M; N, [\theta_m] | M; N, [\theta_m] \rangle = 1, \tag{19}$$

iff $W(M, N)$ is given by

$$W(M, N) = \frac{(M + N - 1)!}{(N - 1)! \, M!} \qquad \text{(bosons)}, \tag{20}$$

$$W(M, N) = \frac{N!}{(N - M)! \, M!} \qquad \text{(fermions)}. \tag{21}$$

The sums over the phases in Eq. (19) are actually trivial since the inner product does not depend on the phases, but we purposely introduce them for the later convenience. Equations (20) and (21) are precisely the degeneracies in Bose-Einstein and Fermi-Dirac statistics, respectively.

Now, following the philosophy mentioned earlier, the interaction is ignored at this stage of shifting to the statistical mechanical description of the system. That is, the vanishing coupling limit,

$$g \rightarrow 0, \tag{22}$$

has to be taken. In this limit, the total energy in Eq. (17) converges to

$$E_{M,N} = M \varepsilon. \tag{23}$$

The phase operators as the observables disappear from the theory, and accordingly $\theta_{m_i}$'s in the quantum states become irrelevant and should be eliminated. This could be somewhat analogous to the concept of random phases [18]. Thus, in view of Eq. (19), the physical state in the vanishing coupling limit is given by the density matrix summed over the phases:

$$\hat{\rho} = \frac{1}{(s+1)^N} \sum_{m_1, m_2, \ldots, m_N = 0}^{s} |M; N, [\theta_m]\rangle \langle M; N, [\theta_m]|, \tag{24}$$

which is calculated to be

$$\hat{\rho} = \frac{1}{W(M,N)} \sum_{P\{n\}} |n_1, n_2, \ldots, n_N\rangle \langle n_1, n_2, \ldots, n_N| \delta_{n_1 + n_2 + \cdots + n_N, M}. \tag{25}$$

The limit $s \to \infty$ can also be taken for the bosons, now (recall that $s$ is always unity for the fermions). This is precisely the microcanonical density matrix of the bosons with Eq. (20) or the fermions with Eq. (21). We ascertain that both perfect decoherence and equal *a priori* probability are simultaneously realized, as desired.

Thus, the microcanonical ensembles are derived for the bosons and fermions from the present scheme in a unified manner.

Equilibrium thermostatistics is formulated in the ordinary way. The value of the von Neumann entropy $S = -\text{tr}(\hat{\rho} \ln \hat{\rho})$ for the state in Eq. (25) yields the Boltzmann relation $S = \ln W(M, N)$ with the Boltzmann constant being set equal to unity. The inverse temperature is calculated by the use of the thermodynamic relation $\beta = \partial S / \partial E_{M,N}$. For large $N$, we have $\beta \cong (1/\varepsilon) \ln(N\varepsilon / E_{M,N} \pm 1)$, where "+" is for the bosons and "−" for the fermions.

Derivation of canonical ensembles is now straightforward. We divide the total isolated system into the objective system $S$ consisting of $N_S$ oscillators and the heat bath $B$ of $N_B$ oscillators, provided that the condition $N_B \gg N_S$ is imposed. $N = N_S + N_B$, $N_S$ and $N_B$ are all fixed. The energy $E_{M,N} \equiv E_{S, M_S, N_S} + E_{B, M_B, N_B} = M\varepsilon$ with $M = M_S + M_B$ ($M_B \gg M_S$) is fixed, but each of $M_S$ and $M_B$ is not fixed. Then, we rewrite the state in Eq. (18) as follows:

$$|M; N, [\theta_m]\rangle = \sum_{M_S, M_B} \sqrt{\frac{W_S(M_S, N_S) W_B(M_B, N_B)}{W(M, N)}} |M_S; N_S, [\theta_{S,m}]\rangle_S$$
$$\otimes |M_B; N_B, [\theta_{B,m}]\rangle_B \, \delta_{M_S + M_B, M}, \qquad (26)$$

where

$$|M_A; N_A, [\theta_{A,m}]\rangle_A = \frac{1}{\sqrt{W_A(M_A, N_A)}} \sum_{P\{n_A\}} |n_{A,1}, n_{A,2}, ..., n_{A,N_A}\rangle_A \, \delta_{n_{A,1} + n_{A,2} + ... + n_{A,N_A}, M_A}$$
$$\times \exp\left(i \sum_{i=1}^{N_A} n_{A,i} \, \theta_{A, m_i}\right) \qquad (A = S, B). \qquad (27)$$

In Eq. (26), the symbol $\Sigma_{M_A}$ ($A = S, B$) stands for the summation over all possible number of excitations. The state in Eq. (25) is then rewritten as

$$\hat{\rho} = \frac{1}{W(M,N)} \sum_{M_S, M_B} \sum_{P\{n_S\}, P\{n_B\}} |n_{S,1}, n_{S,2}, ..., n_{S,N_S}\rangle_{S\ S}\langle n_{S,1}, n_{S,2}, ..., n_{S,N_S}|$$

$$\otimes |n_{B,1}, n_{B,2}, ..., n_{B,N_B}\rangle_{B\ B}\langle n_{B,1}, n_{B,2}, ..., n_{B,N_B}|$$

$$\times \delta_{n_{S,1}+n_{S,2}+\cdots+n_{S,N_S}, M_S}\ \delta_{n_{B,1}+n_{B,2}+\cdots+n_{B,N_B}, M_B}\ \delta_{M_S+M_B, M}\ . \qquad (28)$$

The canonical density matrix of $S$ is obtained from this by the partial trace over $B$, leading to

$$\hat{\rho}_S = \frac{1}{W(M,N)} \sum_{M_S} W_B(M-M_S, N_B)$$

$$\times \sum_{P\{n_S\}} |n_{S,1}, n_{S,2}, ..., n_{S,N_S}\rangle_{S\ S}\langle n_{S,1}, n_{S,2}, ..., n_{S,N_S}| \delta_{n_{S,1}+n_{S,2}+\cdots+n_{S,N_S}, M_S}\ . \qquad (29)$$

From the normalization condition $\mathrm{tr}_S\hat{\rho}_S = 1$, it follows that $W(M,N) = \sum_{M_S} W_S(M_S, N_S) W_B(M-M_S, N_B)$. Since $M_S \ll M$, $W_B(M-M_S, N_B) \cong \exp(-\beta E_{S, M_S, N_S}) W_B(M, N_B)$ with $\beta$ being the inverse temperature defined in terms of the entropy of the heat bath $S_B = \ln W_B(M_B, N_B)$ as $\beta = \partial S_B / \partial E_{B, M_B, N_B}$, where $E_{A, M_A, N_A} = M_A \varepsilon$ ($A = S, B$).

In conclusion, we have constructed the interaction Hamiltonians for bosons and fermions based on the gauge-theoretical structure associated with the phase operators and have seen how entanglement of the specific types is induced by the interactions. Then, we have derived both Bose-Einstein and Fermi-Dirac statistics from the stationary Schrödinger equation in the vanishing interaction limit.

**Acknowledgments**

YY would like to thank Huaqiao University for hospitality extended to him. The work of SA has been supported in part by a grant from National Natural Science Foundation of China (No. 11775084) and Grant-in-Aid for Scientific Research from the Japan Society for the Promotion of Science (No. 26400391 and No. 16K05484), and by the Program of Competitive Growth of Kazan Federal University from the Ministry of Education and Science of the Russian Federation.